\documentclass{PoS}
\usepackage{bm}

\title{Unpolarised TMD PDFs and FFs and the role of transverse momentum dependence
in azimuthal spin asymmetries}

\ShortTitle{Unpolarised TMDs and spin asymmetries}

\author{M.~Anselmino$^{a,b}$, M.~Boglione$^{a,b}$, U.~D'Alesio$^{c,d}$, \speaker{F.~Murgia}$^{\,\,d}$ and A.~Prokudin$^{e,f}$\\
 \llap{$^a$} Dipartimento di Fisica, Universit\`a di Torino, Via P.~Giuria 1, I-10125, Torino, Italy\\
 \llap{$^b$} INFN, Sezione di Torino, Via P.~Giuria 1, I-10125, Torino, Italy\\
 \llap{$^c$} Dipartimento di Fisica, Universit\`a di Cagliari, Cittadella Universitaria, I-09042 Monserrato (CA), Italy\\
 \llap{$^d$} INFN, Sezione di Cagliari, Cittadella Universitaria, I-09042 Monserrato (CA), Italy\\
 \llap{$^e$} Science Division, Penn State University Berks, Reading, Pennsilvania 19210 USA\\
 \llap{$^f$} Theory Center, Jefferson Lab, 12000 Jefferson Avenue, Newport News, Virginia 23606, Usa\\
 Email: \email{mauro.anselmino@to.infn.it}, \email{elena.boglione@to.infn.it},
 \email{umberto.dalesio@ca.infn.it}, \email{francesco.murgia@ca.infn.it}, \email{prokudin@jlab.org}}

\abstract{In the TMD approach, the average transverse momentum of the unpolarised TMD PDFs and FFs
is crucial not only to reproduce unpolarised cross sections and hadron multiplicities,
but also for the understanding of
azimuthal and spin asymmetries. Information on these transverse momenta
is nowadays obtained mainly by fitting multiplicities data for SIDIS, where the intrinsic
motion in the initial parton distributions and in the hadronisation process are strongly correlated
and difficult to estimate separately without ambiguities.
In this contribution we discuss the consequences of this correlation effects on the predictions
for the Sivers and Collins asymmetries measured in SIDIS and $e^+e^-$ annihilations, and under active
investigation for Drell-Yan processes at RHIC and at CERN by the COMPASS experiment.
We show that these effects may be relevant and can sensibly modify the size of the predicted asymmetries.
Therefore, they must be taken into careful account when investigating other aspects of TMDs, like the evolution properties
of the Sivers and Collins functions and the expected process dependence of the Sivers function.}

\FullConference{23rd International Spin Physics Symposium - SPIN2018 -\\
		10-14 September, 2018\\
		Ferrara, Italy}

\begin{document}

\section{Introduction}
The transverse momentum dependent (TMD) formalism is nowadays the most accredited
theoretical approach aiming at explaining a wealth of interesting and puzzling experimental results,
collected over the last years, on single spin and azimuthal asymmetries in semi-inclusive
deep inelastic scattering (SIDIS), Drell-Yan, and $e^+e^-\to h_1 h_2 + X$ processes
(see e.g.~Ref.~\cite{DAlesio:2007bjf} for an introduction to the subject).
In this approach, a new class of transverse momentum dependent parton distributions
(TMD PDFs) and fragmentation functions (TMD FFs), also known collectively as TMDs for short,
are ultimately responsible for the azimuthal asymmetries measured at the hadronic level.
At leading twist, for a spin 1/2 hadron like the proton, there are 8 independent TMD PDFs.
Three of them survive in the collinear configuration and correspond to the well-known
unpolarised, longitudinally polarised, and transversely polarised quark
parton distribution functions. In the fragmentation sector, for (pseudo)scalar particles, like
pions and kaons, there are at leading twist only 2 independent TMDs,
the unpolarised fragmentation function (surviving in the collinear configuration) and the Collins FF.

Among the TMDs, besides the unpolarised functions, the Sivers
distribution~\cite{Sivers:1989cc} and the Collins FF~\cite{Collins:1992kk}
are of special relevance. In fact, they can be responsible, alone or in combination, for many of the most interesting
spin and azimuthal asymmetries observed.
We will limit our considerations to these two TMDs in the sequel.


A crucial point in the TMD approach is the phenomenological parametrisation of the explicit
transverse momentum dependence of TMD PDFs and FFs. To this end, a simple Gaussian shape,
or combinations of Gaussian shapes multiplied by appropriate powers of the transverse momentum,
are commonly adopted. For unpolarised TMDs, the most relevant parameter is the average square
transverse momentum, $\langle k_\perp^2\rangle$ and $\langle p_\perp^2\rangle$, respectively
for PDFs and FFs. The choice of these parameters affects subsequent predictions for the spin
and azimuthal asymmetries.

At present most of the information on the unpolarised TMDs, and on the Sivers and
Collins functions and the related spin and azimuthal asymmetries, comes from (un)polarised
data for SIDIS processes,  $\ell p \to \ell^\prime h + X$. However, in this case
the transverse momentum dependences of the
TMD  PDFs and FFs are strongly correlated. In fact, the trasverse momentum of the observed
hadron, $\bm{P}_{T}$, is kinematically related to a combination of $\bm{k}_\perp$ and  $\bm{p}_\perp$.
In particular, at leading order in a $k_\perp/Q$ power expansion (here $k_\perp$ is used generically
for any intrinsic transverse motion and $Q$ for the large energy scale in the process),
$\bm{P}_{T}\simeq \bm{p}_\perp + z\,\bm{k}_\perp$, where $z$ is the light-cone momentum fraction in the fragmentation process.
As a consequence of this strong correlation, comparably good fits of SIDIS data can be
obtained with \emph{even very} different combinations of the parameters $\langle k_\perp^2\rangle$
and $\langle p_\perp^2\rangle$. However, these comparable fits to SIDIS data may lead to
sensibly different predictions of single spin and azimuthal asymmetries when used in
Drell-Yan processes, where only $\bm{k}_\perp$ effects in the initial PDFs are present,
or, alternatively, in $e^+e^-\to h_1 h_2 + X$ processes, where only $\bm{p}_\perp$
effects in the fragmentation sector play a role. In the rest of this contribution we will
summarize a detailed analysis aiming at clarifying these aspects of TMD phenomenology.

\section{Theoretical approach}
In order to avoid unessential complications and single out the main qualitative results of interest
here, we work in a simplified scheme. We adopt a TMD factorization approach at leading order and leading twist,
considering factorised longitudinal and transverse momentum dependences in all (un)polarised TMDs involved.
The longitudinal component is taken proportional to the corresponding unpolarised collinear function times
a further $x$ (or $z$) dependent term, while for the $k_\perp$ (or $p_\perp$) dependent component we
adopt flavour-independent, Gaussian (or Gaussian-like) functional forms, respecting known positivity bounds.
Moreover, all (un)polarised cross sections (the numerators and denominators of the spin and azimuthal asymmetries)
are integrated over the transverse momentum of the observed hadron(s) (for SIDIS and $e^+e^-$ processes),
or of the lepton pair (for Drell-Yan processes) in the range of validity of the TMD approach,
$P_T, q_T \simeq 1 - 2 \;\rm{GeV}\,\ll Q$, where $Q$ refers generically to the large energy scale involved in the process.
Notice that from the mathematical point of view, all integrations over $k_\perp$ and $p_\perp$
are performed in the full range $[0,+\infty)$. Due to the Gaussian shapes adopted, however, the main contribution  to the
integrals is confined inside the above region of phenomenological interest and validity of the TMD approach.

In this scheme, cross sections and spin/azimuthal asymmetries factorize into a collinear term and a simple,
transverse momentum integrated, component.
All details and intermediate steps can be found in Ref.~\cite{Anselmino:2018psi}.
Here we limit ourselves to show the essential ingredients and the final results.

The unpolarised TMD PDFs and FFs are parameterised as follows:
\begin{equation}
f_{q/p}(x,k_\perp) = f_{q/p}(x)\,\frac{e^{-k_\perp^2/\langle k_\perp^2\rangle}}{\pi\langle k_\perp^2\rangle}\,,
\qquad D_{h/q}(z,p_\perp) = D_{h/q}(z)\,\frac{e^{-p_\perp^2/\langle p_\perp^2\rangle}}
{\pi\langle p_\perp^2\rangle}\,.
\label{eq:unp-tmds}
\end{equation}
The quark transversity distribution, the Sivers function and the Collins fragmentation functions are
analogously parameterised in the following way:
\begin{equation}
h_1^q(x,k_{\perp}) = h_1^q(x)\,\frac{e^{-k_\perp^2/\langle k_{\perp}^2\rangle_T}}{\pi\langle k_{\perp}^2\rangle_T}\,,
\label{eq:h1-tmd}
\end{equation}
\begin{equation}
\Delta^N f_{q/p^\uparrow} (x,k_\perp) = \Delta^N f_{q/p^\uparrow}(x)\;
\sqrt{2e}\,\frac{k_\perp}{M_S} \; e^{-k_\perp^2/M^2_S}\,
\frac{e^{-k_\perp^2/\langle k_\perp^2\rangle}}{\pi\langle k_\perp^2\rangle}
\equiv \Delta^N f_{q/p^\uparrow}(x) \; \sqrt{2e}\,\frac{k_\perp}{M_S} \;
\frac{e^{-k_\perp^2/\langle k_\perp^2\rangle_S}}{\pi\langle k_\perp^2\rangle}\,,
\label{eq:siv}
\end{equation}
\begin{equation}
\Delta^N  D_{h/q^\uparrow}(z,p_\perp) = \Delta^N  D_{h/q^\uparrow}(z)\;
\sqrt{2e}\,\frac{p_\perp}{M_C} \; e^{-p_\perp^2/M_C^2}\,
\frac{e^{-p_\perp^2/\langle p_\perp^2\rangle}}{\pi\langle p_\perp^2\rangle}
\equiv \Delta^N  D_{h/q^\uparrow}(z)\;
\sqrt{2e}\,\frac{p_\perp}{M_C} \;
\frac{e^{-p_\perp^2/\langle p_\perp^2\rangle_C}}{\pi\langle p_\perp^2\rangle}\,,
\label{eq:coll}
\end{equation}
where
\begin{equation}
\langle k_\perp^2\rangle_S = \frac{\langle k_\perp^2\rangle\, M_S^2}
{\langle k_\perp^2\rangle + M_S^2}\,\qquad
\langle p_\perp^2\rangle_C = \frac{\langle p_\perp^2\rangle\, M_C^2}
{\langle p_\perp^2\rangle + M_C^2}\,.
\label{eq:kspc}
\end{equation}

Of interest here are the transverse momentum integrated expressions of the Sivers
asymmetry (for SIDIS and Drell-Yan processes) and of the Collins asymmetry (for SIDIS and
$e^+e^-\to h_1h_2 + X$ processes). Again, we report below the final results. All details can be found
in Ref.~\cite{Anselmino:2018psi}.

\subsection{\label{sec:sidis-siv} $P_T-$integrated Sivers asymmetry for the SIDIS process $\ell p^\uparrow
\to \ell^\prime h + X$}

\begin{equation}
A_{UT}^{\sin(\phi_h-\phi_S)}(x,z) =
A^S_{\rm DIS}(x,z)\,{\cal F}^S_{\rm DIS}(z)\,,
\quad {\rm with}\quad
A^S_{\rm DIS}(x,z) = \frac{\sum_q\,e_q^2\,\Delta^N f_{q/p^\uparrow}(x)\,D_{h/q}(z)}
{2\,\sum_q\,e_q^2\,f_{q/p}(x)\,D_{h/q}(z)}\,,
\label{eq:A-UT-AF-int-s}
\end{equation}
\begin{equation}
{\cal F}^S_{\rm DIS}(z,\rho_S,\xi_1) = \sqrt{\frac{e\pi}{2}}\,\left[\,
\frac{\rho_S^3(1-\rho_S)}{\rho_S + \xi_{1}/z^2}\right]^{1/2}\,,
\quad\quad
\xi_1 = \frac{\langle p_\perp^2\rangle}{\langle k_\perp^2\rangle}\,,\quad\quad
\rho_S = \frac{\langle k_\perp^2\rangle_S}{\langle k_\perp^2\rangle} =
\frac{1}{1+\frac{\langle k_\perp^2\rangle}{M_S^2}}\,.
\label{eq:S-sidis}
\end{equation}

\subsection{\label{sec:collins-siv} $P_T-$integrated Collins asymmetry for the SIDIS process $\ell p^\uparrow
\to \ell^\prime h + X$}

\begin{equation}
A_{UT}^{\sin(\phi_h+\phi_S)}(x,y,z) = A^C_{\rm DIS}(x,y,z)
\,{\cal F}^C_{\rm DIS}(z)\,,
\quad A^C_{\rm DIS}(x,y,z) = \frac{1-y}{1+(1-y)^2} \,
\frac{\sum_q\,e_q^2\,h_1^q(x)\,\Delta^N D_{h/q^\uparrow}(z)}
{\sum_q\,e_q^2\,f_{q/p}(x)\,D_{h/q}(z)}\,,
\label{eq:A-UT-AF-coll-int-s}
\end{equation}
\begin{equation}
{\cal F}^C_{\rm DIS}(z,\rho_C,\xi_1/\xi_T) = \sqrt{\frac{e\pi}{2}}\,\left[\,
\frac{\rho_C^3(1-\rho_C)}{\rho_C + z^2(\xi_{T}/\xi_1)}\right]^{1/2}\,,
\quad
\xi_T = \frac{\langle k_\perp^2\rangle_T}{\langle k_\perp^2\rangle}\,, \quad
\rho_C = \frac{\langle p_\perp^2\rangle_C}{\langle p_\perp^2\rangle} =
 \frac{1}{1+\frac{\langle p_\perp^2\rangle}{M_C^2}}\,.
\label{eq:C-sidis}
\end{equation}

\subsection{\label{sec:sidis-DY} $P_T-$integrated Sivers asymmetry for the Drell-Yan process
$h_1^\uparrow h_2\to \ell^-\ell^+ + X$}

\begin{equation}
A_N^{\rm DY}(y,M) = A^S_{\rm DY}(x_1,x_2)\,{\cal F}^S_{\rm DY}\,,
\quad A^S_{\rm DY}(x_1,x_2) \equiv A^S_{\rm DY}(y,M) =
\frac{\sum_q e_q^2\,\Delta^N f_{q/h_1^\uparrow}(x_1) f_{\bar q/h_2}(x_2)}
{2\sum_q e_q^2\,f_{q/h_1}(x_1) f_{\bar q/h_2}(x_2)}\,,
\label{eq:A-DY-AF-int-s}
\end{equation}
\begin{equation}
{\cal F}^S_{\rm DY}(\rho_S,\xi_{21}) = \sqrt{\frac{e\pi}{2}}\,\left[\,
\frac{\rho_S^3(1-\rho_S)}{\rho_S+\xi_{21}}\right]^{1/2}\,,
\quad\quad
\xi_{21} = \frac{\langle k_{\perp 2}^2\rangle}{\langle k_{\perp 1}^2\rangle}\,,
\quad\quad
\rho_S = \frac{\langle k_\perp^2\rangle_S}{\langle k_{\perp 1}^2\rangle} =
\frac{1}{1+\frac{\langle k_{\perp 1}^2\rangle}{M_S^2}}\,.
\label{eq:S-DY}
\end{equation}

\subsection{\label{sec:collins-ee} $P_T-$integrated Collins asymmetry for $e^+e^-\to h_1 h_2 + X$ (hadronic-plane frame)}

\begin{equation}
P_0^{h_1h_2}(z_1,z_2;\theta) =
A_{\rm ee}^{h_1h_2}(z_1,z_2;\theta)\,{\cal F}_{\rm ee}^C\,,
\quad\quad
{\cal F}_{\rm ee}^C(\rho_C) = 2\,e\,\rho_C^2(1-\rho_C)\,,
\label{eq:p0-af-ee-int-s}
\end{equation}
\begin{equation}
A_{\rm ee}^{h_1h_2}(z_1,z_2;\theta) = \frac{1}{4}\,
\frac{\sin^2\theta}{1+\cos^2\theta}\,
\frac{z_1z_2}{z_1^2+z_2^2}\,
\frac{\sum_q e_q^2\,\Delta^N D_{h_1/q^\uparrow}(z_1)\,
\Delta^N D_{h_2/\bar q^\uparrow}(z_2)}
{\sum_q e_q^2\,D_{h_1/q}(z_1)\,D_{h_2/\bar q}(z_2)}\,.
\label{eq:C-ee}
\end{equation}

Notice that for simplicity we are assuming here that the hadrons $h_{1,2}$ are both pions or kaons.
Cases like $\pi K$ pairs would in general require two different values of $\langle p_\perp^2\rangle$.
Similar results can be obtained adopting the thrust-axis frame.

\section{\label{sec:pheno} Phenomenological results}

In order to clarify our discussion and reduce the number of free parameters,
we will perform some additional simplifying assumptions: a) Concerning the transversity
distribution, we will assume that $\langle k_\perp^2\rangle_T  = \langle k_\perp^2\rangle$,
that is $\xi_T \equiv 1$; b) We only consider Drell-Yan processes in proton-proton collisions;
 this amounts to take $\langle k_{\perp 1}^2\rangle = \langle k_{\perp 2}^2\rangle
\equiv \langle k_{\perp}^2\rangle$, and $\xi_{21}\equiv 1$ in the sequel;
The inclusion of the COMPASS case, for DY in $\pi p$ collisions, would require two
independent values for the average square transverse momentum in the initial pion and proton
beams.

In this simplified but realistic scheme, the transverse momentum integrated components
of the asymmetries (the functions ${\cal F}^{S,C}$ in Eqs.~(\ref{eq:A-UT-AF-int-s})-(\ref{eq:C-ee}))
depend only on three parameters:
1) $\xi_1 = \langle p_\perp^2\rangle/\langle k_\perp^2\rangle$, estimated by fitting
unpolarised observables (multiplicities, the Cahn effect); 2)  $\rho_{S,C}$, that are
fixed, using $\xi_1$, by fitting available data on spin/azimuthal asymmetries.
Notice that in the SIDIS case there is a residual dependence on $z$, the light-cone
momentum fraction in the fragmentation process.

A few comments on these crucial parameters are in order here.
From the mathematical point of view, $0 < \xi_1 < + \infty$, the lower (upper) limit corresponding
respectively to a completely collinear configuration in the fragmentation (distribution) sector.
Physically, these limiting values are quite extreme and highly unrealistic.
A phenomenologically plausible range of values is $0.15 < \xi_1 < 2.2$
(see e.g.~Fig.~9 of Ref.~\cite{Bacchetta:2017gcc}).
Concerning the parameters $\rho_{S,C}$, they govern, in our scheme, the transverse momentum
dependence of the Sivers and Collins functions w.r.t.~that of the corresponding unpolarised
functions. Mathematically $0 < \rho_{S,C} < 1$, however also in this case the limiting values
are phenomenologically unrealistic.

Now the crucial point of our analysis is the following: How do our predictions on the Sivers and
Collins asymmetries in SIDIS, Drell-Yan and $e^+e^-$ annihilation processes depend on the choice
of specific values for $\xi_1$ and consequently $\rho_{S,C}$?

The most general way to answer this question is to study how the transverse components of the
asymmetries, the ${\cal F}^{S,C}$ functions, change when we move along a generic trajectory
in the $(\xi_1,\rho)$ parameter space, starting from some fixed point $(\hat{\xi}_1,\hat{\rho})$,
corresponding to a phenomenological reference fit of unpolarised observables,
and subject to possible constraints dictated by the available experimental data on spin/azimuthal asymmetries.
In particular, we want to study how these changes reflect on the collinear components of the
Sivers and Collins functions and, ultimately, on our predictions for the corresponding asymmetries.

\subsection{ The Sivers asymmetry in SIDIS and Drell-Yan processes}

Let us to this end concentrate first on the Sivers case, and consider, to be definite, two
different and comparably good parameterisations of the quark Sivers functions delivered by the Cagliari-Torino
group:
\begin{enumerate}
\item The fit of Ref.~\cite{Anselmino:2008sga}, referred as FIT09 in the sequel, for which
\begin{equation}
\langle k_\perp^2\rangle = 0.25\, {\rm GeV}^2,\qquad \langle p_\perp^2\rangle
= 0.20\, {\rm GeV}^2,
\qquad M_S^2 = 0.34\, {\rm GeV}^2\,,
\label{eq:fit09-1}
\end{equation}
that implies
\begin{equation}
\hat{\xi}_1^{(09)} = 0.80,\qquad \hat{\rho}_S^{(09)} = 0.58\,.
\label{eq:fit09-2}
\end{equation}
\item The fit of Ref.~\cite{Anselmino:2016uie}, referred as FIT16 in the sequel, for which
\begin{equation}
\langle k_\perp^2\rangle = 0.57\, {\rm GeV}^2,\qquad \langle p_\perp^2\rangle
= 0.12\, {\rm GeV}^2,
\qquad M_S^2 = 0.80\, {\rm GeV}^2\,,
\label{eq:fit16-1}
\end{equation}
implying
\begin{equation}
\hat{\xi}_1^{(16)} = 0.21,\qquad \hat{\rho}_S^{(16)} = 0.58\,.
\label{eq:fit16-2}
\end{equation}
\end{enumerate}

All details on the fitting procedures adopted  and on the extraction of all parameters
(including those for the collinear component of the quark Sivers functions, not mentioned here)
can be found in the quoted references.

Notice that we can always reformulate the general expressions of the functions ${\cal F}^{S,C}(\xi_1,\rho)$
in the $(\xi_1,\rho)$ parameter space through a rescaling factor times their value in the fixed starting point,
${\hat{\cal F}}^{S,C}(\hat{\xi_1},\hat{\rho})$. More in detail, for the asymmetries of interest, we can write:

\begin{equation}
{\cal F}^{S}_{\rm DIS}(z,\xi_1,\rho_S) = R^S_{\rm DIS}\,\hat{{\cal F}}^{S}_{\rm DIS}(z,\hat{\xi}_1,\hat{\rho}_S)\,,
\qquad R^{S}_{\rm DIS} = \left[\,\frac{\rho_S^3(1-\rho_S)}{\rho_S+\xi_1/z^2}
\,\frac{\hat{\rho}_S+\hat{\xi}_1/z^2}{\hat{\rho}_S^3(1-\hat{\rho}_S)}\,
\right]^{1/2}
\label{eq:fs-dis-r}
\end{equation}

\begin{equation}
{\cal F}^S_{\rm DY}(\xi_{21}=1,\rho_S) = R^S_{\rm DY}\,\hat{{\cal F}}^{S}_{\rm DY}(\hat{\rho}_S)\,,
\qquad R^{S}_{\rm DY} = \left[\,\frac{\rho_S^3(1-\rho_S)}{\rho_S+1}
\,\frac{\hat{\rho}_S+1}{\hat{\rho}_S^3(1-\hat{\rho}_S)}\,\right]^{1/2}
\label{eq:fs-dy-r}
\end{equation}

\begin{equation}
{\cal F}^{C}_{\rm DIS}(z,\xi_1,\rho_C) = R^C_{\rm DIS}\,\hat{{\cal F}}^{C}_{\rm DIS}(z,\hat{\xi}_1,\hat{\rho}_C)\,,
\qquad R^C_{\rm DIS} = \left[\,\frac{\rho_C^3(1-\rho_C)}{\rho_C+z^2/\xi_1}\,
\frac{\hat{\rho}_C+z^2/\hat{\xi}_1}{\hat{\rho}_C^3(1-\hat{\rho}_C)}\,\right]^{1/2}
\end{equation}

\begin{equation}
{\cal F}^{C}_{\rm ee}(\rho_C) = R^C_{\rm ee}\,\hat{{\cal F}}^{C}_{\rm ee}(\hat{\rho}_C)\,,
\qquad R^C_{\rm ee} = \frac{\rho_C^2(1-\rho_C)}{\hat{\rho}_C^2(1-\hat{\rho}_C)}\,.
\end{equation}

Let us now consider in more detail the Sivers case. Almost all data available
come from SIDIS, mainly from the HERMES and COMPASS experiments.
Therefore, we will assume that all acceptable trajectories in the $(\xi_1,\rho_S)$ parameter
space are bound to preserve the value of the total Sivers asymmetry:
\begin{equation}
A^S_{\rm DIS}(x,z)\,{\cal F}^S_{\rm DIS}(z,\xi_1,\rho_S) \simeq
\hat{A}^S_{\rm DIS}(x,z)\,\hat{{\cal F}}^S_{\rm DIS}(z,\hat{\xi}_1,\hat{\rho}_S)\,,
\label{eq:asfs-fix}
\end{equation}
or, equivalently,
\begin{equation}
{\cal F}^S_{\rm DIS} = R^{S}_{\rm DIS}\, \hat{{\cal F}}^S_{\rm DIS}\,,
\quad\quad A^S_{\rm DIS} \simeq \frac{1}{R^{S}_{\rm DIS}}\,\hat{A}^S_{\rm DIS}\quad\quad {\rm with} \quad\quad
R^{S}_{\rm DIS} = \left[\,\frac{\rho_S^3(1-\rho_S)}{\rho_S+\xi_1/z^2}
\,\frac{\hat{\rho}_S+\hat{\xi}_1/z^2}{\hat{\rho}_S^3(1-\hat{\rho}_S)}\,
\right]^{1/2}\,,
\label{eq:RFDIS}
\end{equation}

On the contrary, since only very few, low-statistic data are presently available for the Sivers asymmetry
in Drell-Yan processes, they are not used in the fits of the Sivers function. As a result, the full DY Sivers
asymmetry is not constrained like the SIDIS one. However, given that both the SIDIS and DY asymmetries depend
linearly on the Sivers function, we may reasonably assume that
\begin{equation}
\frac{A^S_{\rm DY}}{\hat{A}^S_{\rm DY}} \simeq
\frac{A^S_{\rm DIS}}{\hat{A}^S_{\rm DIS}}\quad \Rightarrow
A^S_{\rm DY} \simeq \Bigl(\,\frac{A^S_{\rm DIS}}{\hat{A}^S_{\rm DIS}}\,\Bigr)\,
\hat{A}^S_{\rm DY} \equiv \frac{1}{R^{S}_{\rm DIS}}\,\hat{A}^S_{\rm DY}\,.
\end{equation}

Bearing in mind Eq.~(\ref{eq:fs-dy-r}), we can then write the full DY asymmetry as
\begin{equation}
A^{\rm DY}_N = A^S_{\rm DY} {\cal F}^S_{\rm DY} \simeq
\left(\frac{R^{S}_{\rm DY}}
{R^{S}_{\rm DIS}}\right)\, \hat{A}^S_{\rm DY}\hat{\cal F}^S_{\rm DY} =
R^N_{\rm DY} \hat{A}^{\rm DY}_N\,,\quad{\rm with} \quad
R^N_{\rm DY} = \left[\,\frac{\rho_S+\xi_1/z^2}{\hat{\rho}_S+\hat{\xi}_1/z^2}
\,\frac{\hat{\rho}_S+1}{\rho_S+1}\,\right]^{1/2}\,.
\label{eq:rndy}
\end{equation}

Now, let us come back to the two reference fits for the quark Sivers functions, FIT09 and FIT16,
see respectively Eqs.~(\ref{eq:fit09-2}) and (\ref{eq:fit16-2}).
As we said, both sets reproduce comparably well the SIDIS Sivers asymmetry data.
Notice however that, while $\hat{\rho}_S^{(09)} \simeq \hat{\rho}_S^{(16)}\simeq \hat{\rho}_S\simeq 0.58$,
due to the strong correlation between $\langle k_\perp^2\rangle$ and $\langle p_\perp^2\rangle$
and the corresponding uncertainties, the values of $\hat{\xi}_1^{(09)}$ and $\hat{\xi}_1^{(16)}$
are sizably different.
{}From Eq.~(\ref{eq:rndy}), we therefore see that the predictions of the two reference sets
for the Sivers asymmetry in Drell-Yan processes are related as follows:
\begin{equation}
A_N^{\rm DY}(\hat{\rho}_S,\hat{\xi}_1^{(16)}) \simeq
\left[\,\frac{\hat{\rho}_S + \hat{\xi}_1^{(16)}/z^2}
{\hat{\rho}_S + \hat{\xi}_1^{(09)}/z^2}
\,\right]^{1/2}\hat{A}^{\rm DY}_N(\hat{\rho}_S,\hat{\xi}_1^{(09)})\,.
\label{eq:ANDY-sc1}
\end{equation}

Using the values for the $\hat{\xi}_1$, $\hat{\rho}_S$ parameters given above for the two sets,
one finds that the rescaling factor in Eq.~(\ref{eq:ANDY-sc1}) varies in the range $[0.52,0.68]$ for $0.1 < z < 0.7$.
Since the SIDIS data utilized in the fits are dominated by the small-$z$ region, we find that
\begin{equation}
A_N^{\rm DY}(\hat{\rho}_S,\hat{\xi}_1^{(16)}) \simeq \frac 12
\hat{A}^{\rm DY}_N(\hat{\rho}_S,\hat{\xi}_1^{(09)})\,.
\label{eq:rn05}
\end{equation}

The main outcome of this study is therefore that, because of the unavoidable
strong correlations between the parameters $\langle k_\perp^2\rangle$ and $\langle p_\perp^2\rangle$,
comparably good fits to the
Sivers asymmetry in SIDIS may lead to very different estimates for the same asymmetries in Drell-Yan
processes. Since the Drell-Yan process is crucial in order to study the process dependence
and the evolution properties of the Sivers function, this effect should be taken into careful account,
before drawing any conclusion concerning these aspects.

The formalism presented in the previous pages can be generalised and applied to different
scenarios for the Sivers effect and to the Collins asymmetry in SIDIS and $e^+e^-$ annihilation processes.

\begin{figure}[]
\includegraphics[width=7.5truecm,angle=0]{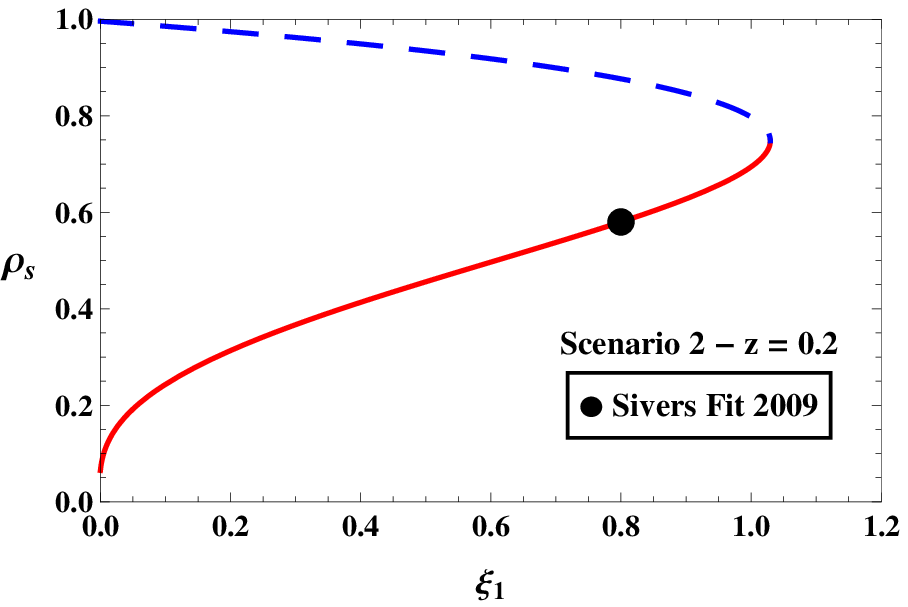}
\includegraphics[width=7.5truecm,angle=0]{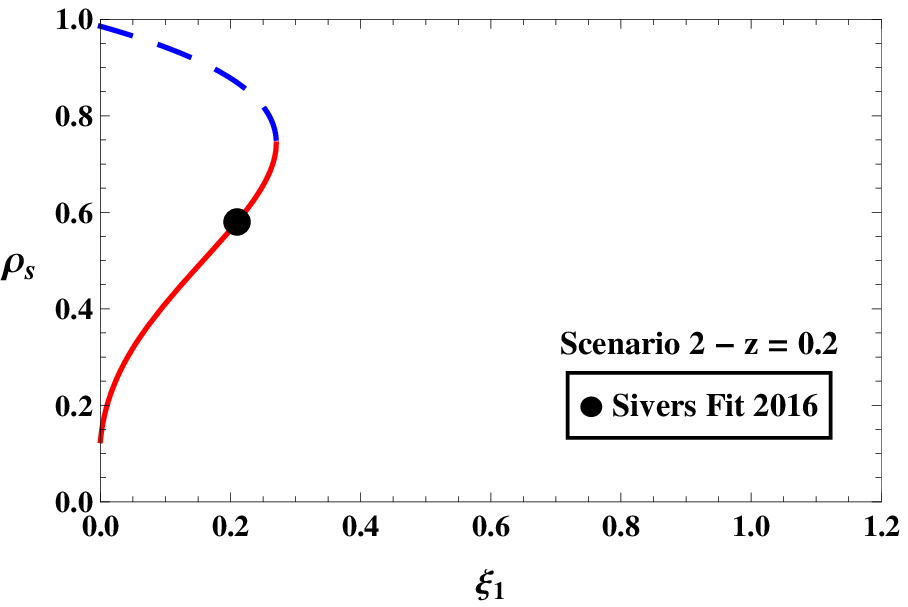}
\caption{Sets of values of $\rho_S$ and $\xi_1$ which leave unchanged ${\cal F}^S_{\rm DIS}(z = 0.2)$,
see Eq.~(\protect\ref{eq:S-sidis}).
The black dots correspond to the fits of Ref.~\protect\cite{Anselmino:2008sga} (left plot, FIT09) and of
Ref.~\protect\cite{Anselmino:2016uie} (right plot, FIT16).
Notice that for each value of $\xi_1$ one finds two possible values of $\rho_S$.
Similar results are found for $z=$ 0.4 or 0.6.}
\label{Fig1}
\end{figure}

\begin{figure}[]
\includegraphics[width=7.5truecm,angle=0]{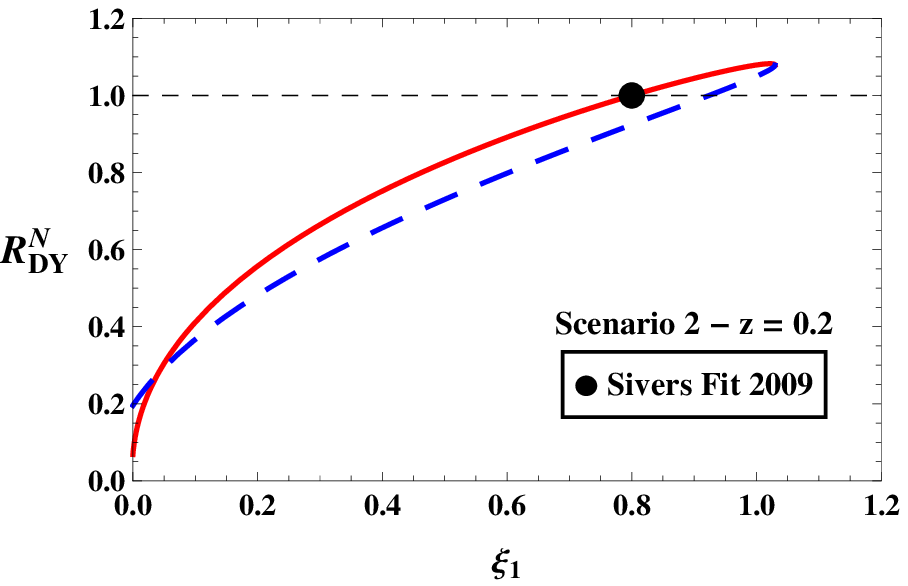}
\includegraphics[width=7.5truecm,angle=0]{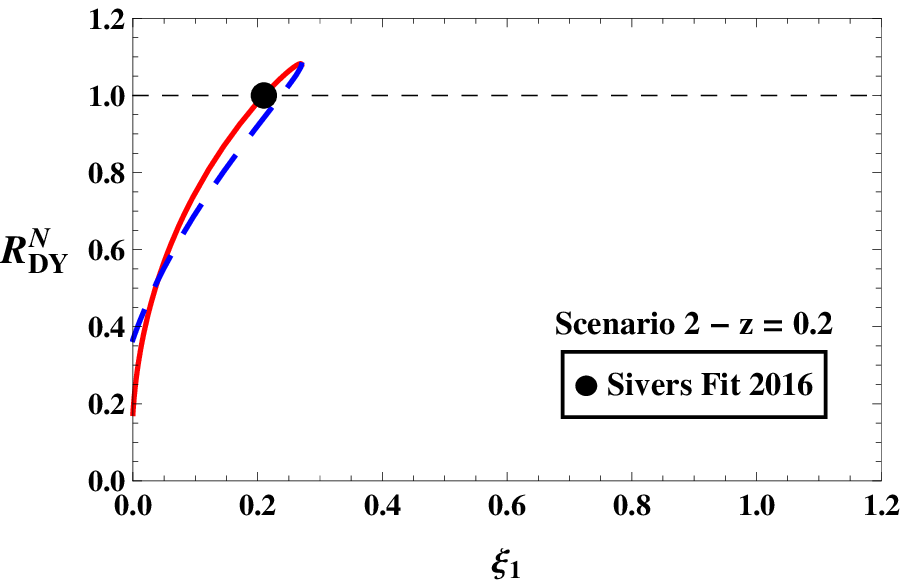}
\caption{Change in the Drell-Yan $q_T$-integrated Sivers asymmetry $A_N^{\rm DY}$,
 Eqs.~(\protect\ref{eq:A-DY-AF-int-s}) and (\protect\ref{eq:S-DY}),
as functions of $\xi_1$, in correspondence of the $\rho_S$ values shown in Fig.~\protect\ref{Fig1}.
The rescaling factor $R^N_{\rm DY}$ is defined in Eq.~(\protect\ref{eq:rndy}).
In this scenario $A_{UT}^{\sin(\phi_h-\phi_S)}$, Eq.~(\protect\ref{eq:A-UT-AF-int-s}), does
not change, together with its components $A^S_{\rm DIS}$ and ${\cal F}^S_{\rm DIS}$.
}
\label{Fig2}
\end{figure}

As an example, instead of fixing the full SIDIS Sivers asymmetry as a whole, when enough data
on the transverse momentum dependence of the asymmetry will be available, one could think of fixing separately
both the collinear ($A^S_{\rm DIS}$) and the $P_T$-integrated (${\cal F}^S_{\rm DIS}$) components of the asymmetry.
{}From Eq.~(\ref{eq:RFDIS}), we see that this corresponds to require $R^S_{\rm DIS}=1$
(the Sivers scenario 2 of Ref.~\cite{Anselmino:2018psi}).
Under this constraint, the allowed values of $\rho_S$  as a function of $\xi_1$ for $z=0.2$, and
the resulting rescaling factor for the Drell-Yan Sivers asymmetry, are shown respectively in Figs.~\ref{Fig1} and
\ref{Fig2}.

Hopefully, in the near future, when more data on the DY Sivers asymmetry will be available from COMPASS and RHIC,
one could try to constrain the allowed values of the $\xi_1$ and $\rho_S$ parameters by requiring
to reproduce both the SIDIS and Drell-Yan Sivers asymmetries at the same time.
These more general scenarios are discussed in depth in Ref.~\cite{Anselmino:2018psi},
to which we refer the reader for further details.

\subsection{ The Collins asymmetry in SIDIS and $e^+e^-$ annihilation processes}

The treatment of the Collins asymmetry is more complicated than the Sivers case.
First of all, the Collins fragmentation function enters linearly in the SIDIS Collins
asymmetry, convoluted with the transversity distribution, while it appears ``quadractically"
(as a convolution of two Collins FFs) in the $e^+e^-$ case.
Therefore, changes in the value of $\xi_1$ can affect only the transversity distribution, or only the
Collins function or, more probably, both of them simultaneously, leaving room to more different possible scenarios.
Moreover, in contrast to the Sivers case,  detailed experimental information is available both for SIDIS and $e^+e^-$ processes.

Also in this case, we have considered two comparable reference fits for the transversity distribution and the Collins FF
with similar values of $\rho_C$ but very different values of $\xi_1$, see Ref.~\cite{Anselmino:2018psi}.
The detailed analysis of several possible scenarios, covering all possibilities mentioned above, shows that
also in the Collins case the estimates of the collinear components of the transversity distribution and
the Collins function may vary as a function of $\xi_1$. However, these changes seem to be milder than in the
Sivers case, apart from some marginal configurations (see Ref.~\cite{Anselmino:2018psi} for more details).

\section{Concluding remarks}
All present parameterisations of the most interesting and most studied TMDs, the quark transversity and Sivers distributions
and the Collins fragmentation function, originate mainly from SIDIS data, to some extent from $e^+e^-$ data and only
marginally from Drell-Yan results.

In this contribution we have investigated, in a simple but general TMD approach, to what extent
the unavoidable strong correlations between the average transverse momenta for the TMD PDFs and FFs,
as estimated from unpolarized SIDIS data, may affect the extraction of the collinear component of the
polarized TMDs and, ultimately, the predictions for spin/azimuthal asymmetries in
Drell-Yan and $e^+e^-\to h_1 h_2 + X$ processes.

We have shown that comparably good fits of the SIDIS Sivers and Collins azimuthal asymmetries
can be obtained with (even very) different values of the ratio
$\xi_1 = \langle p_\perp^2\rangle/\langle k_\perp^2\rangle$. As a consequence,
the corresponding estimates for the Sivers asymmetry in Drell-Yan processes can differ by a factor
of up to 2.

Concerning the extraction of the Collins fragmentation function and of the transversity distribution from
SIDIS and $e^+e^-$ annihilation data, the uncertainty on $\xi_1$ seems to have milder (but still not negligible)
effects, except for some marginal cases.

This analysis shows that a more precise knowledge of $\langle k_\perp^2\rangle$,
$\langle p_\perp^2\rangle$ and $\xi_1$ is crucial. This is particularly true since we are entering
a new stage in the exploration of the 3D structure of hadrons, aiming at a more precise
determination of the functional shapes of the TMD PDFs and FFs, the understanding of their process dependence and a full
implementation of TMD evolution.
For this, new experimental results from RHIC, Jlab, COMPASS and the future
planned Electron Ion Collider will be crucial.

\providecommand{\href}[2]{#2}\begingroup\raggedright\endgroup


\end{document}